\documentclass[runningheads]{llncs}

\usepackage{epsfig}
\usepackage{graphicx}
\usepackage{amsmath}
\usepackage{amssymb}
\usepackage{mathtools}
\usepackage{multirow}
\usepackage{authblk}
\usepackage[symbol]{footmisc}
\usepackage{multirow}
\usepackage{ctable} 
\usepackage{subfig}
\usepackage{floatrow}
\usepackage{dashbox}
\usepackage{colortbl}
\usepackage{xcolor}
\usepackage{comment}
\usepackage{color}
\usepackage{bm}
\usepackage{arydshln}

\def\bal#1\eal{\begin{align}#1\end{align}} 

\newcommand {\bbmtx}{\begin{bmatrix}} 
\newcommand {\ebmtx}{\end{bmatrix}} 

\begin{document}


\title{Reconstructing the Noise Manifold for Image Denoising}

\titlerunning{Abbreviated paper title}

\author{Ioannis Marras\inst{1} \and
Grigorios G. Chrysos\inst{3} \and
Ioannis Alexiou\inst{1} \and
Gregory Slabaugh\inst{1,2} \and
Stefanos Zafeiriou\inst{3}}

\authorrunning{I. Marras et al.}

\institute{Huawei Noah's Ark \and
University College London \and Imperial College London}

\maketitle

\begin{abstract}

Deep Convolutional Neural Networks (CNNs) have been successfully used in many low-level vision problems like image denoising. Although the conditional image generation techniques have led to large improvements in this task, there has been little effort in providing conditional generative adversarial networks (cGAN)~\cite{mirza2014conditional} with an explicit way of understanding the image noise for object-independent denoising reliable for real-world applications. The task of leveraging structures in the target space is unstable due to the complexity of patterns in natural scenes, so the presence of unnatural artifacts or over-smoothed image areas cannot be avoided. To fill the gap, in this work we introduce the idea of a cGAN which explicitly leverages structure in the image noise space. By learning directly a low dimensional manifold of the image noise, the generator promotes the removal from the noisy image only that information which spans this manifold. This idea brings many advantages while it can be appended at the end of any denoiser to significantly improve its performance. Based on our experiments, our model substantially outperforms existing state-of-the-art architectures, resulting in denoised images with less oversmoothing and better detail.

\end{abstract}

\section{Introduction}

During image acquisition, due to the presence of noise some image corruption is inevitable and can degrade the visual quality considerably. Therefore, removing noise from the acquired image is a key step for many computer vision and image analysis applications~\cite{gonzalez2002digital}. As an indispensable step in many digital imaging and computer vision systems, image denoising has been investigated for decades, while it is still an active research topic.

\begin{figure}[htp]
\begin{center}
  \includegraphics[width=0.5\columnwidth]{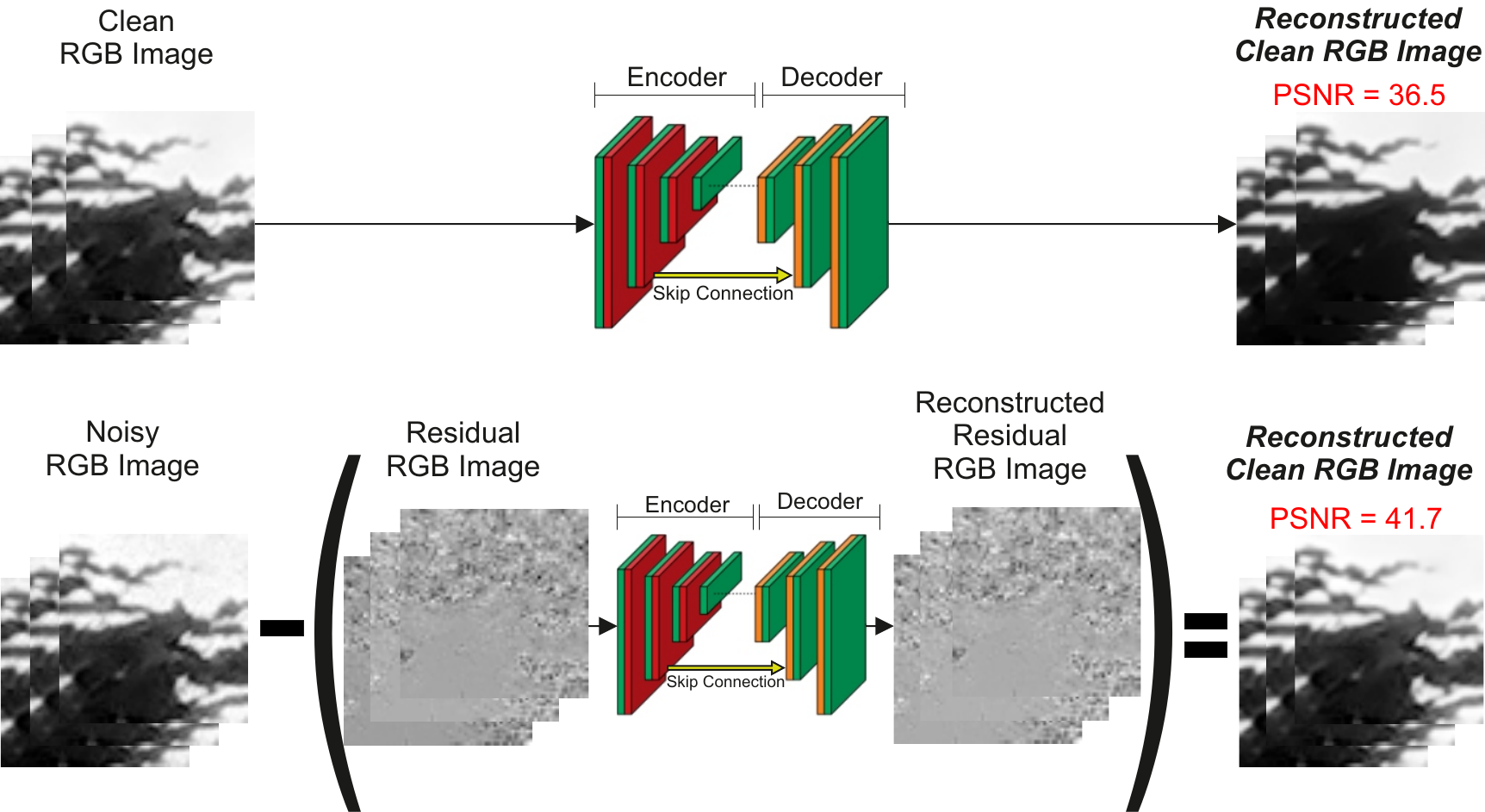}\vspace{-0.6cm}
  \caption{\textbf{Motivation of our method}: By characterizing directly the image spatially variant noise, the reconstruction of the clean image is much more accurate. Instead of constraining the output of a generator to span the target space, is better to constrain it to remove from the noisy image only that information which spans the manifold of the residual image. \vspace{-0.6cm}}
  \label{fig:Motivation}
\end{center}
\end{figure}

Denoising algorithms can be grouped in two categories: learning-based and model-based. Modelling the image prior from a set of noisy and ground-truth image sets is the goal of discriminative learning. The performance of the current learning models is limited  by the inadequacy of the current methods to handle all possible levels of noise in a single model. In this category are methods such as brute force learning like MLP~\cite{Burger2012MLP}, CNNs~\cite{zhang2017beyond,zhang2017IRCNN} or truncated inference~\cite{chen2017TNRD}. On the other hand, the model-based algorithms are computationally expensive, unable to characterize complex image textures. In the this category fall algorithms including external priors~\cite{anwar2017category}, Markov random field models~\cite{roth2009fields,tappen2007learning}, gradient methods~\cite{xu2007Iterative,weiss2007makes}, non-local self-similarity~\cite{Dabov2009BM3DSAPCA} and sparsity (e.g.\ MCWNNM~\cite{Gu2014WNN}).

A denoising algorithm should be efficient, perform denoising using a single model and handle spatially variant noise when the noise standard-deviation is known or unknown. The physics of digital sensors and the steps of an imaging pipeline are well-understood and can be leveraged to generate training data from almost any image using only basic information about the target camera sensor. Recent work has shifted to sophisticated signal-dependent single source noise models~\cite{Healey1994RadiometricCC} that better match the physics of image formation~\cite{Liu2008AutomaticEA,Mildenhall_2018_CVPR,Brooks_CVPR2019}. Also, because different camera sensors exhibit different noise characteristics, adapting a learned denoising algorithm to a new camera sensor may require capturing a new dataset. However, capturing noisy and noise-free image pairs is difficult, requiring long exposures or large bursts of images, and post-processing to combat camera motion and lighting changes.

In this paper, we introduce the idea of a cGAN which directly constrains the image spatially variant noise for image denoising (Fig.~\ref{fig:Motivation}). In this way, we avoid the direct characterization of the space of clean images, since the complexity of natural image patterns is extremely high. To do so, a combination of supervised (regression) and unsupervised (autoencoder) `\emph{encoder-decoder}' type subnets applies implicit constraints in the residual image (the difference between the noisy observation and the clean image) latent subspace. By adopting the idea of residual learning~\cite{zhang2017beyond} in the regression subnet and using a shared decoder, the unsupervised subnet is explicitly constrained to generate residual image samples that span only the image noise manifold. Intuitively, this can be thought of as constraining the regression subnet of our dense regression to \emph{subtract from the noisy image only the residual image that looks like realistic image noise coming from a specific camera sensor}. The proposed idea: a) allows the direct association of one or more camera sensors with their corresponding noise statistics and b) introduces also the idea of a discriminator operating directly in the residual image domain. Our system: a) increases significantly the robustness of the image denoising task, b) makes easier the model adaptation to a new camera sensor, c) allows multi-camera noise reduction during one inference step, d) allows multi-source noise removal during one inference step, e) utilizes all the samples in the residual image domain even in the absence of the corresponding noisy input samples, f) can be applied at the end of any residual learning based denoiser improving its performance and g) deals with a wide range of noise levels.

\section{Related Work}

\subsection{Image Prior Based Methods}

\emph{Image prior} based methods, e.g.\ NSCR~\cite{Dong2013NonlocallyCS}, TWSC~\cite{xu2018TWSC},  WNNM~\cite{Gu2014WNN}, can be employed to solve the denoising problem of unknown noise because they do not require training data since they model the image prior over the noisy image directly. The classic internal statistics based method, BM3D~\cite{dabov2007image}, is based on the idea that natural images usually contain repeated patterns. It combines the non-local self-similarity model and sparse model. In \emph{non-local means} (NLM)~\cite{buades2005non}, the pixel values are predicted based on their noisy surroundings. Many variants of NLM and BM3D seeking self-similar patches in different transform domains were proposed, e.g. SAPCA~\cite{Dabov2009BM3DSAPCA}, NLB~\cite{Lebrun2013NLB}. Sparsity is enforced by dictionary-based methods~\cite{Dong2011CSR} by employing self-similar patches and learning over-complete dictionaries from clean images. In contrast, Noise2Void (N2V)~\cite{Krull2018Noise2Void} and Noise2Noise (N2N)~\cite{N2N_2018} do not require training noisy image pairs, nor clean target images. N2N attempts to learn a mapping between pairs of independently degraded versions of the same training image. For image patch restoration, maximum likelihood algorithms like Gaussian Mixture Models (GMMs), were employed to learn statistical priors from image patch groups~\cite{Chen2015External,Xu2015PGGMM}. Dictionary learning based and basis-pursuit based algorithms such as KSVD~\cite{KSVD_2006}, Fields-of-Experts or TNRD~\cite{TNRD_2015} operated by finding image representations where sparsity holds or statistical regularities are well-modeled~\cite{Zoran2011EPLL}. In \cite{NC_2015}, an extension of non-local Bayes approach, named NC, was proposed to model the noise of each patch group to be zero-mean correlated and Gaussian distributed. The disadvantage of this category of methods is that external information from possible many other images taken under the same condition with the image to be denoised cannot be used. Also in many cases, the computational cost at inference is very high. Furthermore, the generalization capabilities are limited because these methods are defined mostly based on human knowledge.

\subsection{Discriminative Deep Learning Methods}

In recent years, CNNs have achieved great success in image denoising. The first attempt of employing CNNs for the regression task of image denoising was made in~\cite{jain2009natural}. \emph{Discriminative deep learning methods} are trained offline, extracting information from ground truth annotated training sets before they are applied to test data. In DnCNN~\cite{zhang2017beyond} and IrCNN~\cite{zhang2017IRCNN} networks, stacked convolution, batch normalization and ReLU layers are used to estimate the residual~\cite{he2016deep} between the noisy input and the corresponding clean image. By adding symmetric skip connections, an improved encoder-decoder network for image denoising based on residual learning was proposed in~\cite{mao2016image}. A densely connected denoising network, named Memnet, constructed in~\cite{tai2017memnet} to enable memory of the network. A multi-level wavelet CNN (MWCNN) model based on a U-Net architecture used in~\cite{MWCNN_2018} to incorporate large receptive field for image denoising. By incorporating non-local operations into a recurrent neural network (RNN), a non-local recurrent network (NLRN) for image restoration presented in~\cite{NLRN_2018}. A network named N$^3$Net~\cite{N3Net_2018} employs the k-nearest neighbor matching in the denoising network to exploit the non-local property of the image features. A fast and flexible network (FFDNet) which can process images with non-uniform noise corruption proposed in~\cite{2017FFDNetTA}. A residual in the residual structure (RIDNet) used in~\cite{Anwar2019RIDNET} to ease the flow of low-frequency information and apply feature attention to exploit the channel dependencies. Recently, a blind denoising model for real photographs named CBDNet~\cite{Guo2019Cbdnet} is composed of two subnetworks: noise estimation and non-blind denoising. It may require manual intervention to improve results. A self-guided network (SGN), which adopts a top-down self-guidance architecture to better exploit image multi-scale information presented in~\cite{SGN_2019_ICCV}. FOCNet network~\cite{Jia2019FOCNetAF} solved a fractional optimal control problem in a multi-scale approach. Although the methods in this category achieved high denoising quality, they cannot work in the absence of paired training data.

\subsection{Generative Models}

GANs were recently trained to synthesize noise~\cite{Chen2018ImageModeling}. Since image noise is generated by the GAN-generator, pairs of corresponding clean and noisy images are obtained for training CNNs. Noise Flow method~\cite{Abdelhamed2019NoiseFN} combined well-established basic parametric noise models (e.g.\ signal-dependent noise) with the flexibility and expressiveness of normalizing flow architectures to model noise distributions observed from large datasets of real noisy images. The flow model is conditioned on critical variables, such as intensity, camera type, and gain settings. However, it was not clear how to quantitatively assess the quality of the generated samples.

\section{Method}
\label{sec:dense_reg_subsp_method}

In this section, we introduce our system for the task of image denoising. The goal is to produce a single clean (RGB or RAW) image from a corresponding single noisy (RGB or RAW) image captured by a handheld camera. Firstly, we give a brief overview of the noise signal in real images (Section~\ref{ssec:im_noise_modeling}). Our method falls in the group of conditional image generation methods, thus to make the paper self-contained we firstly briefly describe this category of methods (Section~\ref{ssec:dense_reg_subsp_rocgan}) before introducing our method (Section~\ref{ssec:dense_our_subsp_rocgan}).

\subsection{Image Noise Modeling In Real-World Images}
\label{ssec:im_noise_modeling}

Camera sensors output RAW data in a linear color space, where pixel measurements are proportional to the number of photoelectrons collected. The primary sources of noise are shot noise, a Poisson process with variance equal to the signal level, and read noise, an approximately Gaussian process caused by a variety of sensor readout effects. The noise is spatially variant (non-Gaussian); hence, the assumption that noise is spatially invariant, employed by many algorithms does not hold for real image noise. These effects are well-modeled by a signal-dependent Gaussian distribution~\cite{Healey1994RadiometricCC}:
\small
\begin{equation}
x_p \sim \mathcal N \left(y_p, \sigma_r^2 + \sigma_s y_p \right) \label{eq:srnoise}
\end{equation}
\normalsize where $x_p$ is a noisy measurement of the true intensity $y_p$ at pixel $p$. The noise parameters $\sigma_r$ and $\sigma_s$: a) are fixed, given a specific camera sensor, for each image and varies from image to image as sensor gain (ISO) changes and b) are different for different camera sensors even for the same ISO value. Since the noise is structured (not random) a low-dimensional manifold for noise exists. A realistic noise model as well as the in-camera image processing pipeline are important aspects in training CNN-based denoising methods for real photographs~\cite{Guo2019Cbdnet,Anwar2019RIDNET}.

\subsection{Conditional Image Generation}
\label{ssec:dense_reg_subsp_rocgan}

In computer vision, the task of conditional image generation is dominated by approaches similar to GAN. The GAN consists of a generator and a discriminator module commonly optimized with alternating gradient descent methods. cGAN extend the formulation by providing the generator with additional labels. In cGAN, the generator $G$ takes the form of an encoder-decoder network where the encoder projects the label into a low-dimensional latent subspace and the decoder performs the opposite mapping.

cGAN and its variants like Robust cGAN~\cite{chrysos2019}, were successfully applied in the past for the task of object-dependent image denoising. Their consist of a generator and the same discriminator. The encoder-decoder generator of Robust cGAN performs a similar regression as its counterpart in cGAN. It accepts a sample from the source domain (noisy image) and maps it to the target domain (clean image) by using a second CNN in the target domain which promotes more realistic regression outputs. Recently in non GAN-based methods, generators adopting similar architecture were proposed for object-dependent image denoising~\cite{WhiteNNer_2019}. Instead of computing the noise by subtracting the predicted signal from the noisy input, a two-tailed CNN is employed – for inferring the clean image and the noise separately. The input noisy image is decoupled to the signal and noise in a latent space, while after both latent representations are fed into a decoder to generate the signal and noise in spatial domain. There are two major drawbacks of all these methods: i) in the absence of skip connections, these methods perform well only in the case of object-dependent image denoising (i.e. face denoising~\cite{chrysos2019}). The need of having different models for different objects makes them unsuitable for digital devices with limited resources (e.g. smartphones) where the run-time performance is of importance. ii) the purpose of their unsupervised learning sub-networks, whose (hidden) layers contain representations of the input data, is to be sufficiently powerful for compressing (and decompressing) the data while losing as little information as possible. However, even in the presence of skip connections, this procedure of defining a nonlinear representation which can accurately reconstruct image patterns from a variety of real complex objects/scenes is not realistic. As a result, these methods very often hallucinate complex image structures by introducing severe blurry effects or unusual image patterns/artifacts.

\subsection{Image denoising based on noise manifold reconstruction}\label{ssec:dense_our_subsp_rocgan}

\begin{figure*}[!h]
    \centering
    \includegraphics[width=1.04\linewidth]{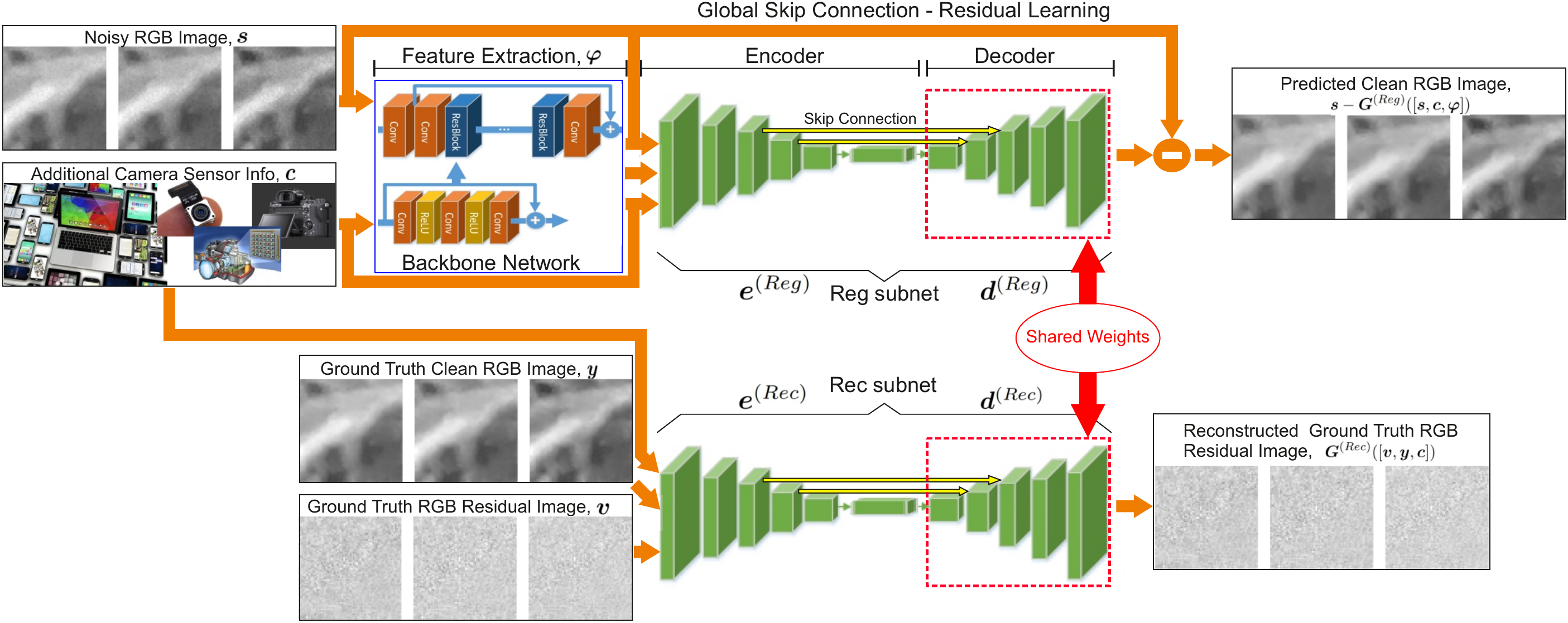}\vspace{-0.3cm}
\caption{Schematic of the proposed generator.}
\label{fig:dense_reg_subsp_ours_schematic}
\end{figure*}

To tackle the problems mentioned in Section \ref{ssec:dense_reg_subsp_rocgan}, the proposed method introduces the general idea of explicitly constraining the residual image removed by a denoiser to lie in the low-dimensional manifold of the image noise source (Section \ref{ssec:im_noise_modeling}). Like cGAN, our method consists of a generator and a discriminator. The generator includes two subnets: the first regression (\emph{Reg}) subnet performs regression while the second reconstruction (\emph{Rec}) is an autoencoder in the residual image domain (unsupervised subnet). Both subnets consist of similar encoder-decoder networks, while a backbone network is used prior to the encoder-decoder network of the \emph{Reg subnet}. By sharing the weights of their decoders, the generator adopts the residual learning strategy to remove from the noisy observation that image information which spans the image noise manifold. A schematic of the proposed generator is illustrated in Fig.~\ref{fig:dense_reg_subsp_ours_schematic}. Rather than directly outputing the denoised image, the supervised \emph{Reg subnet} is designed to predict the ground-truth residual image $\bm{v}=\bm{s}-\bm{y}$, where $\bm{s}$ and $\bm{y}$ stand for the noisy and the clean (ground-truth) image, respectively. Because of that, the unsupervised \emph{Rec subnet} works as a conditional auto-encoder in the domain of $\bm{v}$. The unsupervised subnet during inference is no longer required, therefore the testing complexity remains the same as in standard cGAN. Two Unet style skip connections from the encoder to the decoder used in both subnets improving the learning of the residual between the features corresponding to the image and to the residual image structures.

The reconstruction of $\bm{v}$, by the \emph{Rec subnet}, is an easier task compared to the reconstruction of $\bm{y}$ as in cGAN. In other words, we can learn how to turn bad images into good images by only looking at the structure of the residual image. This property: i) makes our method an object-independent image denoiser and ii) helps the denoiser largely avoid image over-smoothing/artifacts, something essential for image denoising.

Noising is a challenging process to be reversed by the few convolutional layers of the encoder in \emph{Reg subnet} especially in an object-independent scenario. This is why a backbone network used to extract complex feature representations, \bm{$\varphi$}, useful to preserve for later the low and high image frequencies. Different state-of-the-art denoisers could be used as backbone networks, thus the proposed idea could be applied at the end of any denoiser constraining its output improving in that way its performance (Section~\ref{sec:Experiments}).

Furthermore, \emph{Rec subnet} enables the utilization of all the samples in the domain of the residual image even in the absence of the corresponding noisy input samples. For example, in the case of a well defined image noise source, like the one described in Section \ref{ssec:im_noise_modeling}, a huge amount of different residual image realizations (e.g.\ for different ISOs) could be generated and used to train that subnet.

Another advantage is the easier adaptation of an existing trained model to a new camera sensor. To do so, only the \emph{Rec subnet} must be retrained from scratch while the \emph{Reg subnet} needs only to be fine-tuned using a small number of paired training samples obtained using the new sensor. In addition, our method can remove more than one noise source during one inference step. To do so, a different noise manifold for different noise source is obtained, thus a different \emph{Rec subnet} per noise source constrains the denoiser in a sequential manner (Fig.~\ref{fig:MultiNoise_and_DnD_Results}(a)).

The task of learning directly the image noise manifold can be greatly benefit by any conditional information, $\bm{c}$, related to the camera sensor noise characteristics. $\bm{c}$ is explicitly given to both \emph{subnets}. The information that $\bm{c}$ represents varies and directly associates a camera sensor with its corresponding noise statistics. If the camera noise model is known, $\bm{c}$ could contain the two noise parameters $\sigma_r$ and $\sigma_s$ (both same for each pixel) as described in Section \ref{ssec:im_noise_modeling}. Also, one of the advantages of our method is that it supports multi-camera noise reduction during one inference step. To do so, $\bm{c}$ could additionally contain one hot vector per pixel defining the camera id used to take each picture, thus one or more noise sources are explicitly associated to the corresponding camera sensor. More specifically, \emph{Reg subnet} gets as input $\bm{c}$ in concatenation (denoted as [$\cdot$]) with $\bm{s}$ and $\bm{\varphi}$ and outputs $\bm{y}-\bm{G}^{(Reg)}([\bm{s},\bm{c},\bm{\varphi}])$, where $\bm{G}^{(Reg)}([\bm{s},\bm{c},\bm{\varphi}])$ is the predicted residual image. The superscript `Reg' abbreviates modules of the \emph{Reg subnet}. Based on eq.~\ref{eq:srnoise}, the noise variance for a pixel $p$ depends, except for the camera sensor-based parameters, on $y_p$. Thus, the input to \emph{Rec subnet} should be $\bm{v}$ in concatenation with $\bm{y}$ and $\bm{c}$. By giving explicitly $\bm{y}$ as additional input, \emph{the task of the \emph{Rec subnet} is not to learn the underlying structure of a huge variety of complex image patterns, but to learn how clean image structures are affected by the presence of structured noise}.

The proposed idea deals with a wide range of noise levels in contrast to standard cGAN or its variants. In more details, according to~\cite{he2016deep}, when the original mapping $\emph{F}(\bm{s})$ (as in cGAN) is more like an identity mapping, the residual mapping will be much easier to optimize. Note that $\bm{s}$ is much more like $\bm{s}-\bm{G}^{(Reg)}([\bm{s},\bm{c},\bm{\varphi}])$ than $\bm{G}^{(Reg)}([\bm{s},\bm{c},\bm{\varphi}])$ (especially when the noise level is low). Thus, $\emph{F}(\bm{s})$ would be closer to an identity mapping than $\bm{G}^{(Reg)}([\bm{s},\bm{c},\bm{\varphi}])$, and the residual learning formulation is more suitable for image denoising~\cite{zhang2017beyond}.

In the case of image denoising in RGB domain, $\bm{s}$ represents 3-channel image based tensors. Regarding $\bm{s}$ in RAW domain, each pixel in a conventional camera (linear Bayer) sensor is covered by a single red, green, or blue color filter, arranged in a 4-channel Bayer pattern (i.e. R-G-G-B). The content loss consists of two terms that compute the per-pixel difference between the predicted clean image, and the clean (ground-truth) image. The two terms are  i) the $\ell_1$ loss between the ground-truth image and the output of the generator, ii) the $\ell_1$ of their gradients; mathematically expressed as:
\vspace{-0.3cm}
\begin{equation}\label{eq:denoising_content_loss}
\resizebox{1.0\linewidth}{!}{$
\begin{split}
    \mathcal{L}_{c} = \lambda_{c} \cdot \sum_{n=1}^{N} ||(\bm{s}^{(n)}-\bm{G}^{(Reg)}([\bm{s}^{(n)},\bm{c}^{(n)},\bm{\varphi}^{(n)}]))-\bm{y}^{(n)}|| + \lambda_{cg} \cdot \sum_{n=1}^{N} ||\nabla (\bm{s}^{(n)}-\bm{G}^{(Reg)}([\bm{s}^{(n)},\bm{c}^{(n)},\bm{\varphi}^{(n)}])) - \nabla \bm{y}^{(n)}||,
\end{split}$
}
\end{equation}
where $\bm{G}^{(Reg)}([\bm{s}^{(n)},\bm{c}^{(n)},\bm{\varphi}^{(n)}]) = \bm{d}^{(Reg)}(\bm{e}^{(Reg)}([\bm{s}^{(n)},\bm{c}^{(n)},\bm{\varphi}^{(n)}]))$, $N$ stands for the total number of training samples, $\bm{e}$ stands for encoder, $\bm{d}$ stands for decoder and $\lambda_{c}$, $\lambda_{cg}=0.5\cdot\lambda_{ae}$ are hyper-parameters to balance the loss terms. The unsupervised \emph{Rec subnet} contributes the following loss term: \vspace{-0.25cm}

\begin{equation}\label{eq:dense_reg_subsp_ae_loss}
\resizebox{0.55\linewidth}{!}{$
    \mathcal{L}_{Rec} = \sum_{n=1}^{N} [f_d(\bm{v}^{(n)}, \bm{G}^{(Rec)} ([\bm{v}^{(n)},\bm{y}^{(n)},\bm{c}^{(n)}]))]$}
\end{equation}
where $\bm{G}^{(Rec)} ([\bm{v}^{(n)},\bm{y}^{(n)},\bm{c}^{(n)}]) = \bm{d}^{(Rec)}(\bm{e}^{(Rec)}([\bm{v}^{(n)},\bm{y}^{(n)},\bm{c}^{(n)}]))$ is the \emph{Rec subnet}'s output, $f_d$ is a divergence metric ($\ell_2$ loss due to the auto-encoder in the noise domain) and the superscript `Rec' abbreviates modules of the \emph{Rec subnet}. 

Despite sharing the weights of the decoders, the latent representations of the two subnets are forced to span the same space. To further reduce the distance of the two representations in the latent space, a latent loss term $\mathcal{L}_{lat}$ is used. This term minimizes the distance between the encoders' outputs, i.e. the two residual noise representations are spatially close (in the subspace spanned by the encoders). The latent loss term is:  
\vspace{-0.25cm}
\begin{equation}\label{eq:dense_reg_subsp_latent_loss}
\resizebox{0.75\linewidth}{!}{$
    \mathcal{L}_{lat} = \sum_{n=1}^{N} ||\bm{e}^{(Reg)} ([\bm{s}^{(n)},\bm{c}^{(n)},\bm{\varphi}^{(n)}]) - \bm{e}^{(Rec)} ([\bm{v}^{(n)},\bm{y}^{(n)},\bm{c}^{(n)}])||.$}
\end{equation}

As a part of the vanilla cGAN, the feature matching loss~\cite{salimans2016improved,isola2016image} enables the network to match the data and the model's distribution faster. The intuition is that to match the high-dimensional distribution of the data with \emph{Reg subnet}, their projections in lower-dimensional spaces are encouraged to be similar. The feature matching loss is:
\vspace{-0.25cm}
\begin{equation}\label{eq:dense_reg_subsp_projection_loss}
\resizebox{0.65\linewidth}{!}{$
    \mathcal{L}_{f} = \sum_{n=1}^{N} ||\pi(\bm{s}^{(n)} - \bm{G}^{(Reg)}([\bm{s}^{(n)},\bm{c}^{(n)},\bm{\varphi}^{(n)}])) - \pi(\bm{y}^{(n)})||.$}
\end{equation}
where $\pi()$ extracts the features from the penultimate layer of the discriminator.

Skip connections can enable deeper layers to capture more abstract representations without the need of memorizing all the information. The lower-level representations (only) are propagated directly to the decoder through the shortcut, which makes it harder to train the longer path~\cite{rasmus2015semi}. This challenge is implicitly tackled by maximizing the variance captured by the longer path representations. The Decov loss term~\cite{cogswell2016reducing} that penalizes the correlations in the representations (of a layer) and thus implicitly encourages the representations to capture diverse and useful information is used. This loss may be applied to a single layer or multiple layers in a network, while for the $j^{th}$ layer this loss is defined as: 
\vspace{-0.3cm}
\begin{equation}\label{eq:dense_reg_subsp_decov_loss}
\resizebox{0.35\linewidth}{!}{$
    \mathcal{L}^{j}_{decov} = \frac{1}{2} (||\bm{C}^{j}||_F^2 - ||diag(\bm{C}^{j})||_2^2),$}
\end{equation}
where $diag()$ computes the diagonal elements of a matrix and $\bm{C}^{j}$ is the covariance matrix of the $j^{th}$ layer's representations. The loss is minimized when the covariance matrix is diagonal, i.e. it imposes a cost to minimize the covariance of hidden units without restricting the diagonal elements that include the variance of the hidden representations.

\vspace{-0.8cm}
\begin{figure*}[htp]
\begin{center}
  \includegraphics[width=1.04\columnwidth]{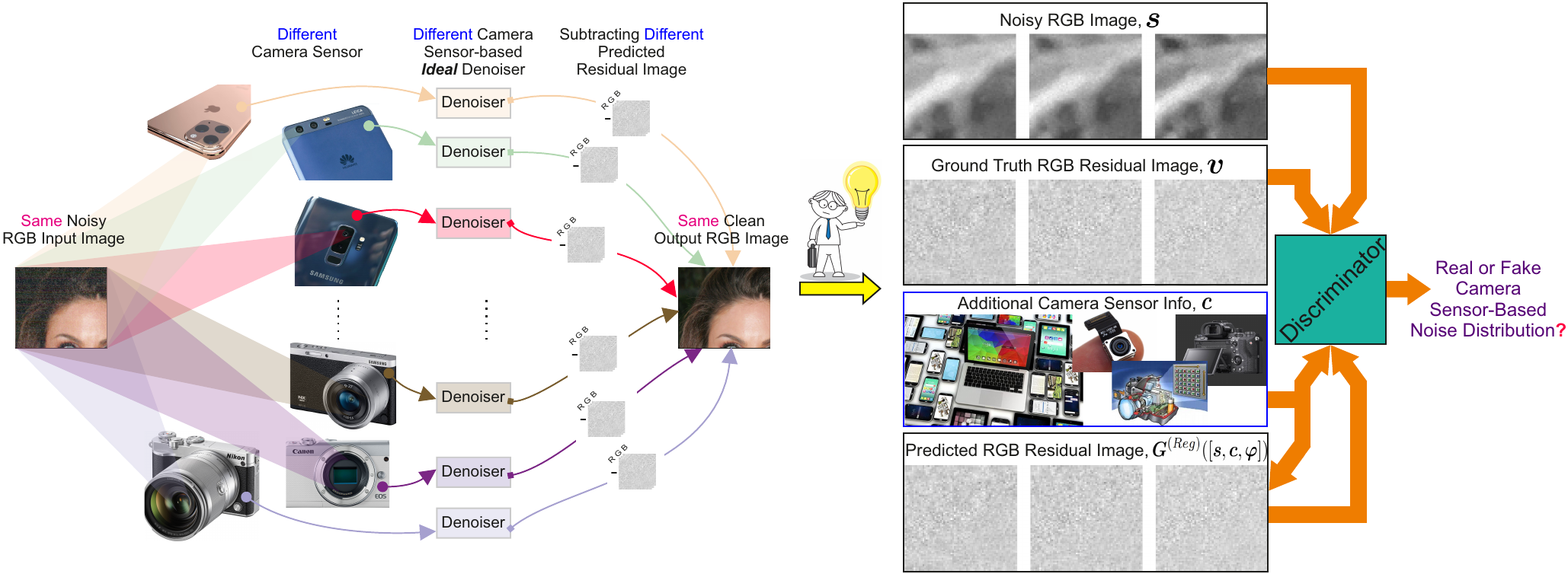}\vspace{-0.6cm}
  \caption{The proposed discriminator which operates directly in the residual image domain. Given the same input image captured by different camera-sensors, the same output image could be obtained if an ideal camera sensor-based denoiser exists although different camera sensors have different noise distributions. \vspace{-0.6cm}}
  \label{fig:Discriminator}
\end{center}
\end{figure*}

In the case of multi-camera noise scenario, let's assume that the same scene is captured under the same lighting conditions by different camera sensors. Let's also assume that an ideal denoiser per camera sensor exists. In that case, the output of all the denoisers should be the same underlying clean image although the noise statistics of each camera can be very different. This leads to the idea of a discriminator which operates directly in the residual image domain (Fig.~\ref{fig:Discriminator}) thus trying to  distinguish between the residual image samples generated from the denoiser and the ground-truth residual image distributions given a specific camera sensor. This is feasible in the proposed method because the \emph{Rec subnet} constraints directly the denoiser to remove only that information which spans the learned noise manifold of each camera sensor. The generator samples $\bm{z}$ from a prior distribution $p_{\bm{z}}$, e.g. uniform, and tries to model the target distribution $p_{d}$; the discriminator $D$ tries to distinguish between the samples generated from the model and the target (ground-truth) image noise distributions. More specifically, the discriminator accepts as input $\bm{G}^{(Reg)}([\bm{s},\bm{c},\bm{\varphi}])$ along with $\bm{v}$, $\bm{c}$ and $\bm{s}$, while the adversarial loss of cGAN (Section \ref{ssec:dense_reg_subsp_rocgan}) is modifying to:
\vspace{-0.2cm}
\begin{equation}\label{eq:dense_reg_subsp_adversarial_loss}
\resizebox{1.0\linewidth}{!}{$
\begin{split}
    \mathcal{L}_{adv}^{\star}(\bm{G}^{(Reg)}, \bm{D}) = \mathbb{E}_{\bm{s},\bm{v} \sim p_{d}(\bm{s},\bm{v})}[\log \bm{D}(\bm{v} | \bm{s},\bm{c})] + 
    \mathbb{E}_{\bm{s} \sim p_{d}(\bm{s}), \bm{z} \sim p_{z}(\bm{z})}[\log (1-\bm{D}(\bm{G}^{(Reg)}([\bm{s},\bm{c},\bm{\varphi}]) | \bm{s},\bm{c}))].
\end{split}$}
\end{equation}
\noindent by solving the following min-max problem:
\vspace{-0.2cm}
\begin{equation}
\resizebox{0.7\linewidth}{!}{$
\begin{split} 
    \min_{\bm{w}_G} \max_{\bm{w}_D} \mathcal{L}_{adv}^{\star}(\bm{G}^{(Reg)}, \bm{D}) = \min_{\bm{w}_G} \max_{\bm{w}_D} \mathbb{E}_{\bm{s},\bm{v} \sim p_{d}(\bm{s},\bm{v})}[\log \bm{D}(\bm{v} | \bm{s}, \bm{c}, \bm{w}_D)] + \\
    \mathbb{E}_{\bm{s} \sim p_{d}(\bm{s}), \bm{z} \sim p_{z}(\bm{z})}[\log (1-\bm{D}(\bm{G}^{(Reg)}([\bm{s},\bm{c},\bm{\varphi}]| \bm{w}_G) | \bm{s},\bm{c},\bm{w}_D))]
    \nonumber
\end{split}$}
\end{equation}
where $\bm{w}_G, \bm{w}_D$ denote the generator's and the discriminator's parameters respectively. The final loss function of our method is:
\vspace{-0.4cm}
\begin{equation}\label{eq:dense_reg_subsp_rocgan_loss}
\resizebox{0.7\linewidth}{!}{$
\begin{split}
    \mathcal{L}_{total} = \mathcal{L}_{adv}^{\star} + \mathcal{L}_{c} + \lambda_{\pi}\cdot\mathcal{L}_{f} + \lambda_{ae}\cdot\mathcal{L}_{Rec} + \lambda_{l}\cdot\mathcal{L}_{lat} + \lambda_{d}\cdot\sum^{j}\mathcal{L}^{j}_{decov},
\end{split}$}
\end{equation}
where $\lambda_{\pi}, \lambda_{ae}$, $\lambda_{l}$ and $\lambda_{d}$ are extra hyper-parameters to balance the loss terms.

\section{Experiments}\label{sec:Experiments}

\subsection{Training Settings}

Synthetic noisy images can be combined with real noisy data to improve the generalization ability of our method to real photographs. To generate them, we follow the pipeline in~\cite{Guo2019Cbdnet} which is based on the noise model (Section~\ref{ssec:im_noise_modeling}). To do so, we employ BSD500~\cite{MartinFTM01}, DIV2K~\cite{Agustsson2017NTIRE2C}, and MIT-Adobe FiveK~\cite{bychkovsky2011learning}, resulting in 3.5K images while for real noisy images, we extract cropped patches of $512\times512$ from SSID~\cite{Abdelhamed2018ACameras} and RENOIR~\cite{anaya2018renoir}. Finally, the data augmentation procedure results in $64\times64$ image patches. In our `\emph{encoder-decoder}' architecture (same for both subnets) 11 layers are used with an latent space of dimensions MB$\times2\times2\times1024$, where MB stands for the mini-batch size. The values of the additional hyper-parameters are $\lambda_{ae}=20$, $\lambda_{l}=0.16$ and $\lambda_{d}=0.9$. The common hyper-parameters $\lambda_{\pi}$ and $\lambda_{c}$ with the vanilla cGAN remain the same. For both subnets: the kernel size used is 3$\times$3; Adam~\cite{kingma2014adam} is used as the optimizer with default parameters; the learning rate is initially set to $10^{-3}$ and then halved after $10^{6}$ iterations; ReLU activation used; the network ran for 50 epochs. 

\subsection{Comparisons on real-World images}
 
The most three challenging public datasets that significantly improve upon earlier (and often unrealistic) benchmarks for denoising, were used to evaluate the performance of our method: the Darmstadt Noise Dataset (DnD)~\cite{Plotz2017BenchmarkingDA}, the Nam Dataset~\cite{Nam2016AHA} and the Smartphone Image Denoising Dataset (SIDD)~\cite{Abdelhamed2018ACameras}. DnD and Nam are multi-camera datasets which allow our method to prove its superiority in performing multi-camera noise reduction. To highlight the contribution of the proposed idea, as the backbone network in our method we used: a) a standard residual (ResNet), introduced in \cite{he2016deep} for image recognition, created by stacking three building blocks, and b) the best deep learning-based method in the literature according to each benchmark, if existing, excluding the last network layer since this network acts as a feature extractor. The pre-trained weights reported in the literature, if available, used as initialization of the backbone network, while it was trained in a end-to-end fashion with the two subnets.

\textbf{Evaluation on DnD}: DnD is a novel benchmark dataset which consists of realistic uncompressed photos from 50 scenes taken by 4 different standard consumer cameras of natural ``in the wild'' scene content. In DnD: the camera metadata has been captured; the noise properties have been carefully calibrated; and the image intensities are presented as RAW unprocessed linear intensities. For each real high-resolution image, the noisy high-ISO image is paired with the corresponding (nearly) noise-free low-ISO ground-truth image.

\newcommand{\DnD}{
\resizebox{0.7\linewidth}{!}{$
\begin{tabular}{ l || l || c|c || c|c || c }
\multicolumn{1}{c||}{} & 
\multicolumn{1}{c||}{} & 
\multicolumn{2}{c||}{\textbf{RAW}} & 
\multicolumn{2}{c||}{\textbf{sRGB}}& \textbf{Runtime} \\
\textbf{Method} & \textbf{Type} & \multicolumn{1}{c|}{\textbf{PSNR}} & \multicolumn{1}{c||}{\textbf{SSIM}} & \multicolumn{1}{c|}{\textbf{PSNR}} & \multicolumn{1}{c||}{\textbf{SSIM}} & (ms) \\
\hline
FoE~\cite{roth2009fields}  & Non-blind &  45.78 &  0.9666 &  35.99 &    0.9042 &   - \\
TNRD~\cite{TNRD_2015} + VST & Non-blind &  45.70 &  0.9609 &    36.09 &   0.8883 &  5{,}200 \\
MLP~\cite{Burger2012MLP} + VST  &  Non-blind & 45.71 &  0.9629 &  36.72 &  0.9122 &  $\sim$60{,}000 \\
MCWNNM~\cite{Gu2014WNN} & Non-blind &  - &  - &  37.38 &  0.9294 &  208{,}100 \\
EPLL~\cite{Zoran2011EPLL} + VST & Non-blind &  46.86 &  0.9730 & 37.46 &  0.9245 &  - \\
KSVD~\cite{KSVD_2006} + VST & Non-blind & 46.87 &  0.9723 &  37.63 &    0.9287 &   $>$60{,}000 \\
WNNM~\cite{Gu2014WNN} + VST & Non-blind & 47.05 &  0.9722 &  37.69 &    0.9260 &   - \\
NCSR~\cite{Dong2013NonlocallyCS} + VST & Non-blind &  47.07 &  0.9688 &  37.79 &    0.9233 &   - \\
BM3D~\cite{dabov2007image} + VST &  \textcolor{purple}{Non-blind} & \textcolor{purple}{47.15} & \textcolor{purple}{0.9737} &  \textcolor{purple}{37.86} & \textcolor{purple}{0.9296} & \textcolor{purple}{6{,}900} \\
Whitenner~\cite{WhiteNNer_2019} & Blind & 47.16 & 0.9737 & 37.88 & 0.9307 & 48 \\
RoCGAN~\cite{chrysos2019} & Blind & 47.17 & 0.9738 & 37.90 & 0.9310 & 49 \\
TWSC~\cite{xu2018TWSC} & Blind &  - &  - &  37.94 &   0.9403 &  195{,}200 \\
CBDNet~\cite{Guo2019Cbdnet} & Blind & - &  - &  38.06 & 0.9421 &  400 \\
DnCNN~\cite{zhang2017beyond} & Blind &  47.37 &   0.9760 &   38.08 &    0.9357  & 60 \\
N$^3$Net~\cite{N3Net_2018} & Blind & 47.56 & 0.9767 & 38.32 & 0.9384 & 210 \\
RIDNet~\cite{Anwar2019RIDNET} & Blind & - &  - &  39.23 & 0.9526 & 215 \\
\textcolor{blue}{UPI}~\cite{Brooks_CVPR2019}  & \textcolor{blue}{Blind} & \textcolor{blue}{48.89} & \textcolor{blue}{0.9824} & \textcolor{blue}{40.35} & \textcolor{blue}{0.9641} & \textcolor{blue}{22} \\
\hdashline 
\textbf{Ours} (empty $\bm{c}$, \textbf{ResNet}~\cite{he2016deep}) & Blind & 49.90 (+1.01) & 0.9861 & 41.50 (+1.15) & 0.9759 & 52 \\
\textbf{Ours} (empty $\bm{c}$, \textcolor{blue}{\textbf{UPI}}) & \textcolor{blue}{Blind} & \textcolor{blue}{50.05} (\textcolor{red}{+1.16}) & \textcolor{blue}{0.9866} & \textcolor{blue}{41.59} (\textcolor{red}{+1.24}) & \textcolor{blue}{0.9760} & \textcolor{blue}{64} \\
\textbf{Ours} (\emph{Non} empty $\bm{c}$, \textbf{ResNet}) & \textcolor{purple}{Non-Blind} & \textcolor{purple}{50.91} (\textcolor{red}{+3.76}) & \textcolor{purple}{0.9873} & \textcolor{purple}{42.11} (\textcolor{red}{+4.25}) & \textcolor{purple}{0.9775} & \textcolor{purple}{63} \\
\toprule
\multicolumn{7}{c}{} \\
\multicolumn{7}{c}{Ablations of \textbf{Ours} (empty $\bm{c}$, \textbf{ResNet}~\cite{he2016deep})} \\
\hline
Standard Discriminator~\cite{mirza2014conditional} & Blind & 49.50 (+0.61) & 0.9835 & 41.0 (+0.65) & 0.9714 & 52 \\
No \emph{Rec subnet} & Blind & 47.51 (-1.38) & 0.9766 & 38.54 (-1.81) & 0.9417 & 52 \\
No \emph{Rec subnet}, No res. learning & Blind & 46.92 (-1.97) & 0.9725 & 37.73 (-2.62) & 0.9316 & 52 \\
\toprule
\end{tabular}$}
}

\newcommand{\MultiNoise}{
  \includegraphics[width=0.3\columnwidth]{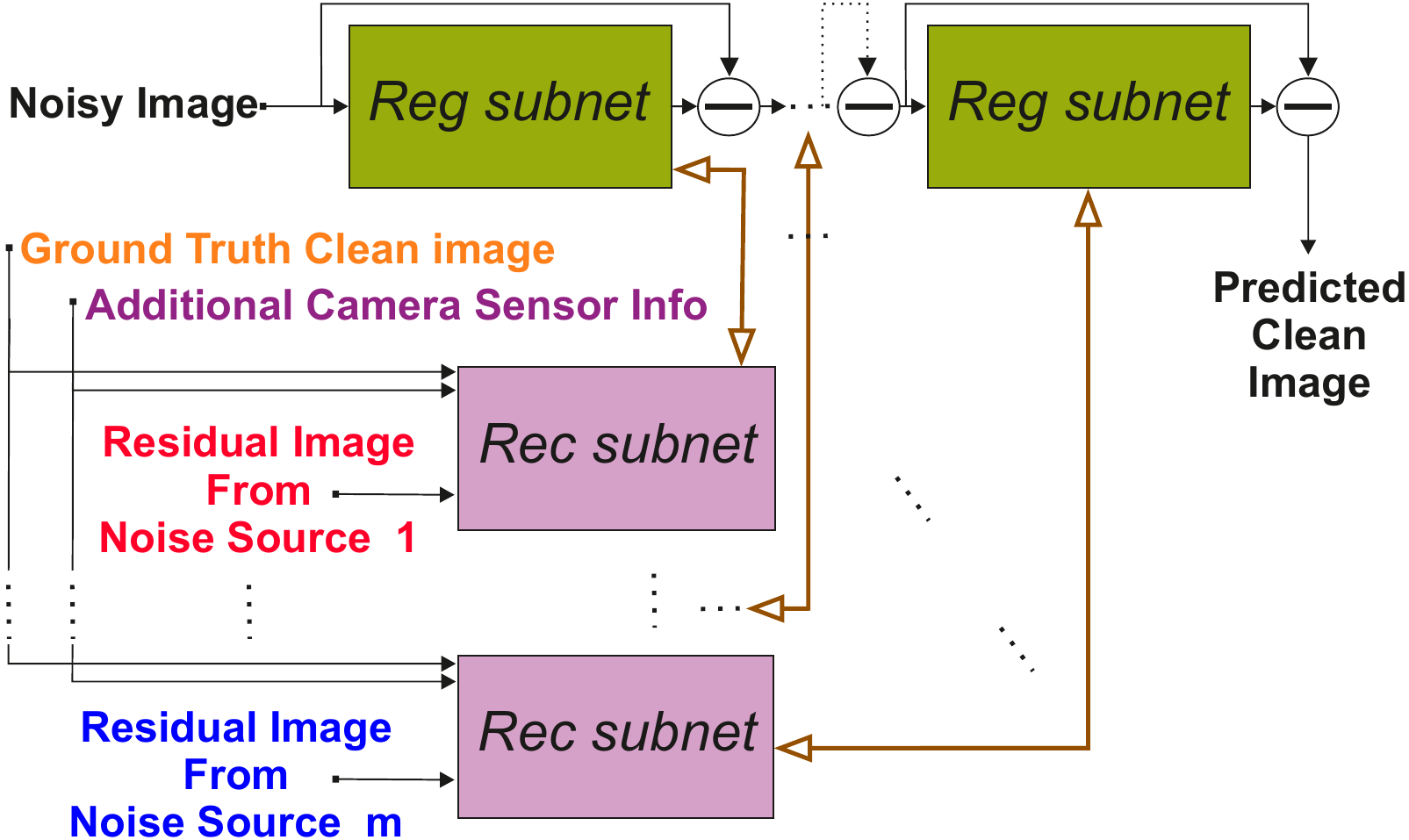}\vspace{-0.6cm}
}

\vspace{-0.9cm}
\begin{figure}%
  \centering
  \subfloat[][]{\MultiNoise}%
  \subfloat[][]{\DnD}\vspace{-0.4cm}
  \caption{(a) In the case of camera multi-source image noise, more than one \emph{Rec subnet} can be employed. Each subnet is responsible for removing noise structure that comes from a specific noise source, and (b) the quantitative results on the DnD benchmark of our method and its ablations. Regarding our method, in parentheses we define the type of denosing plus the used backbone network.\label{fig:MultiNoise_and_DnD_Results}}%
\end{figure}

\begin{figure*}[!t]
\setlength{\abovecaptionskip}{0.2cm}
\centering
\subfloat{
\begin{minipage}[t]{0.2\textwidth}
\centering
\includegraphics[width=1\textwidth]{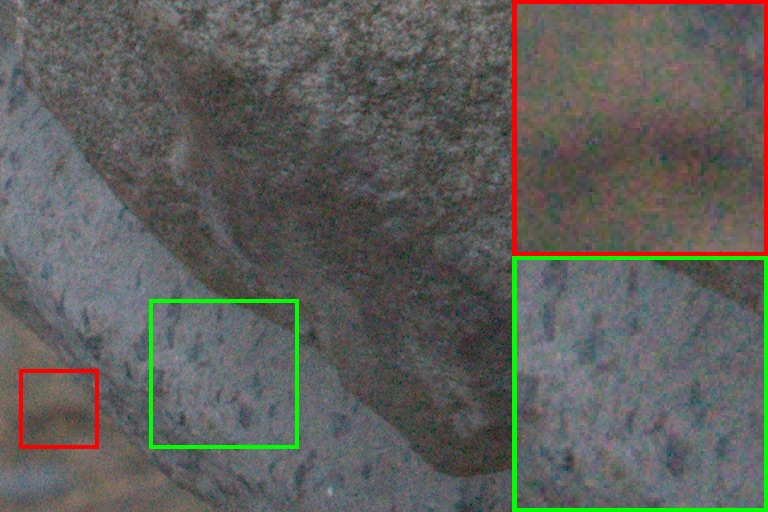}
{\footnotesize  (a) \resizebox{0.55\linewidth}{!}{Noisy image}}
\end{minipage}\hspace{-0.07cm}

\begin{minipage}[t]{0.2\textwidth}
\centering
\includegraphics[width=1\textwidth]{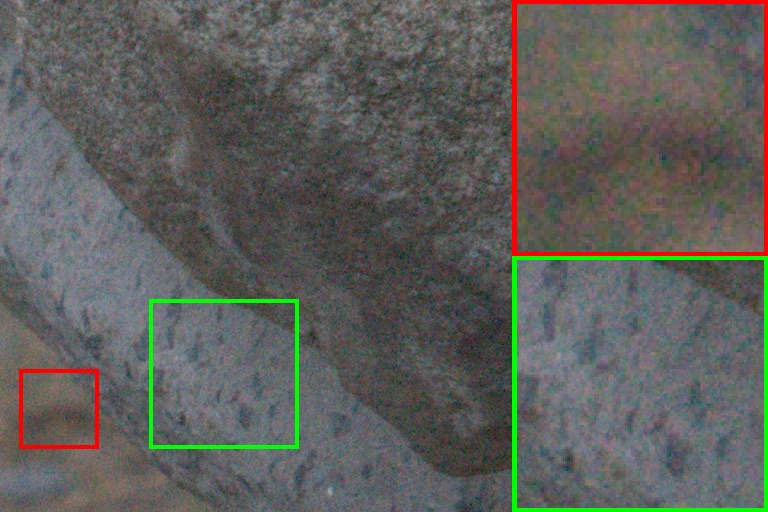}
{\footnotesize  (b) \resizebox{0.45\linewidth}{!}{BM3D~\cite{dabov2007image}}}
\end{minipage}\hspace{-0.07cm}

\begin{minipage}[t]{0.2\textwidth}
\centering
\includegraphics[width=1\textwidth]{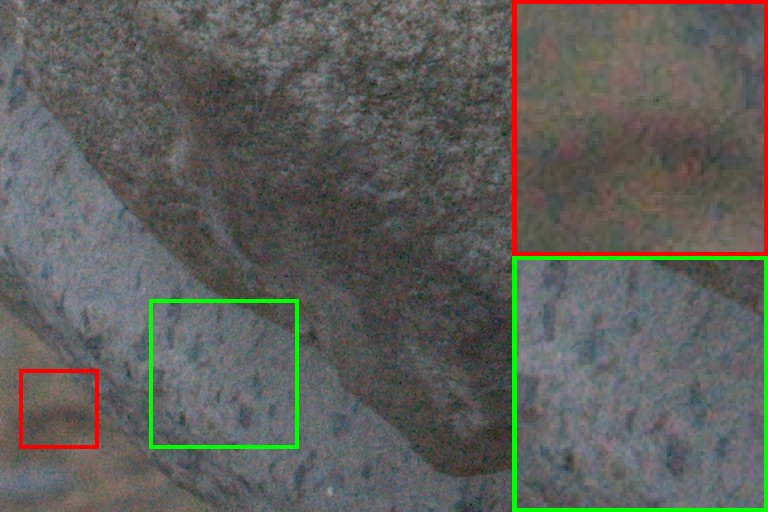}
{\footnotesize  (c) \resizebox{0.55\linewidth}{!}{CDnCNN-B~\cite{zhang2017beyond}}}
\end{minipage}\hspace{-0.07cm}

\begin{minipage}[t]{0.2\textwidth}
\centering
\includegraphics[width=1\textwidth]{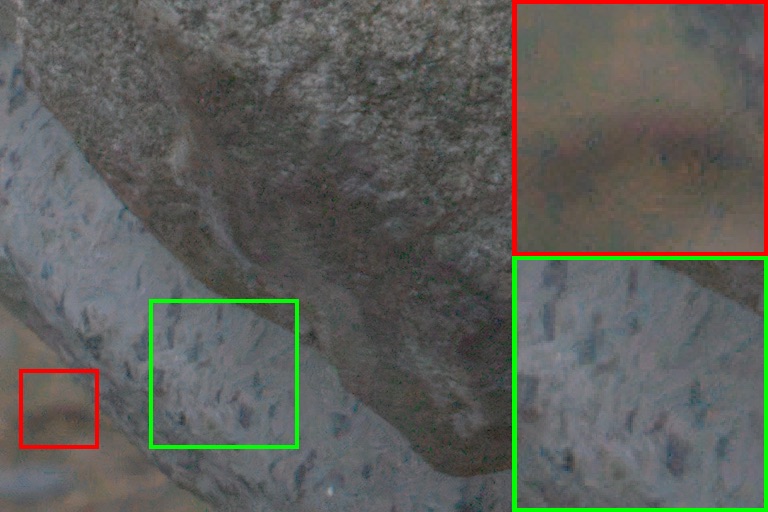}
{\footnotesize  (d) \resizebox{0.35\linewidth}{!}{NC~\cite{NC_2015}}}
\end{minipage}\hspace{-0.07cm}

\begin{minipage}[t]{0.2\textwidth}
\centering
\includegraphics[width=1\textwidth]{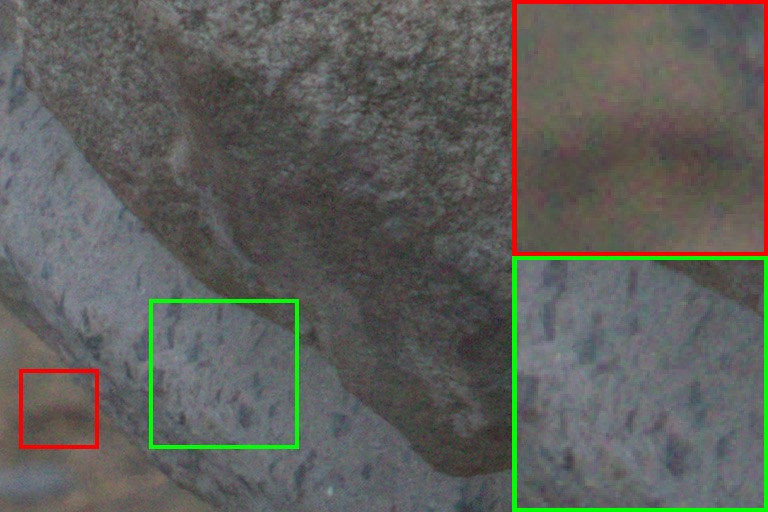}
{\footnotesize  (e) \resizebox{0.55\linewidth}{!}{MCWNNM~\cite{Gu2014WNN}}}
\end{minipage}\hspace{-0.07cm}

}

\subfloat{

\begin{minipage}[t]{0.2\textwidth}
\centering
\includegraphics[width=1\textwidth]{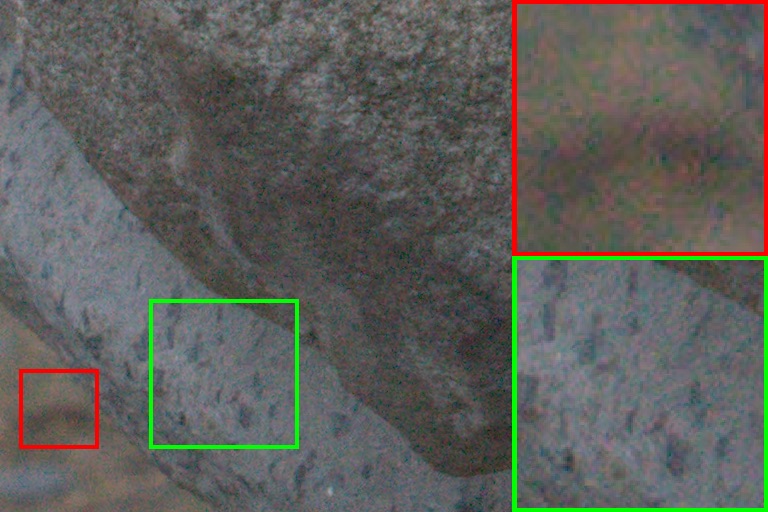}
{\footnotesize  (f) \resizebox{0.5\linewidth}{!}{TWSC~\cite{xu2018TWSC}}}
\end{minipage}\hspace{-0.07cm}

\begin{minipage}[t]{0.2\textwidth}
\centering
\includegraphics[width=1\textwidth]{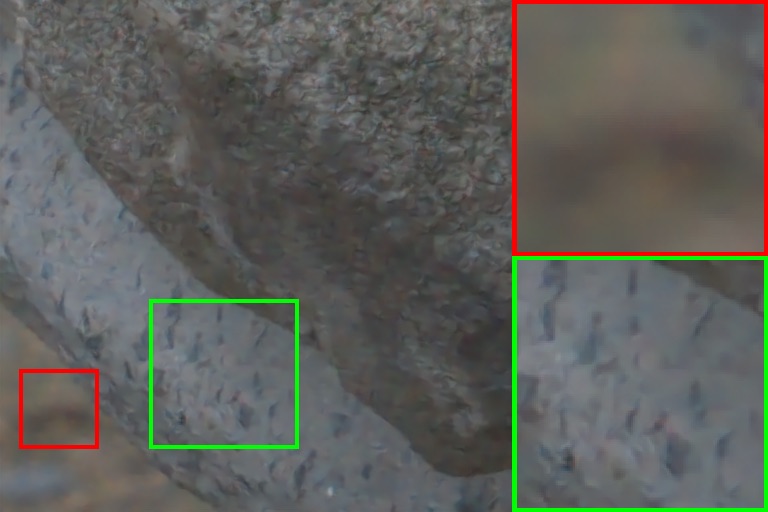}
{\footnotesize  (g) \resizebox{0.5\linewidth}{!}{CBDNet~\cite{Guo2019Cbdnet}}}
\end{minipage}\hspace{-0.07cm}

\subfloat{
\begin{minipage}[t]{0.2\textwidth}
\centering
\includegraphics[width=1\textwidth]{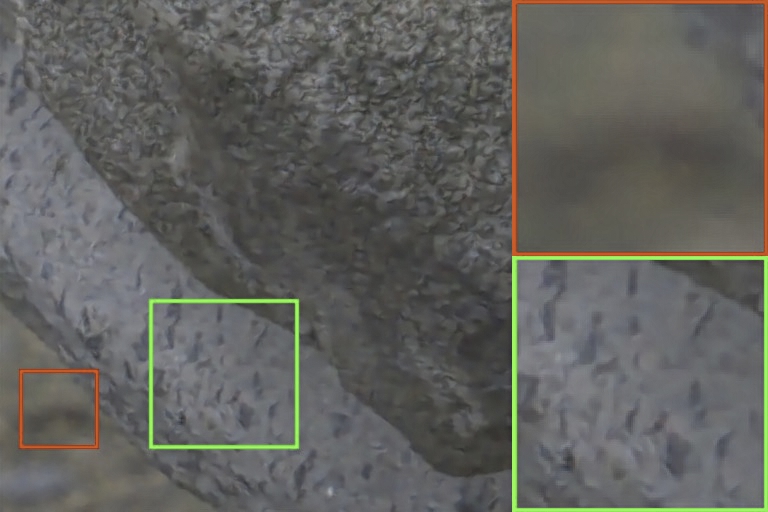}
{\footnotesize  (h) \resizebox{0.5\linewidth}{!}{\textbf{Ours} (Blind)}}    \end{minipage}\hspace{-0.07cm}
}


\begin{minipage}[t]{0.2\textwidth}
\centering
\includegraphics[width=1\textwidth]{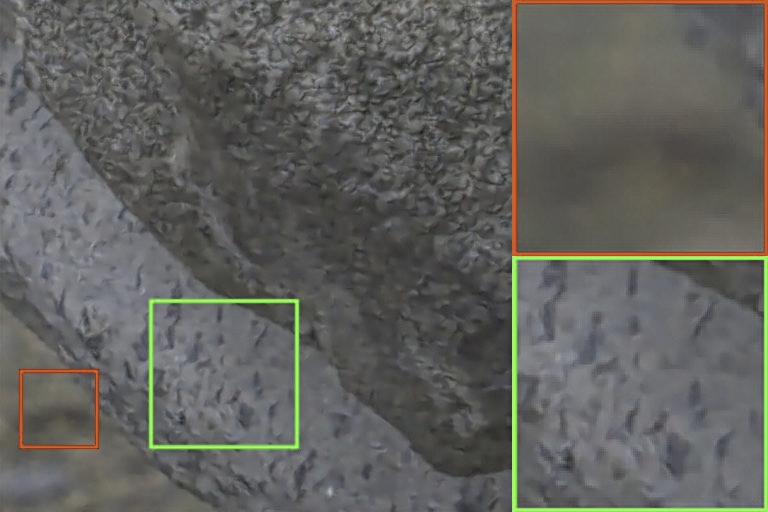}
{\footnotesize  (g) \resizebox{0.7\linewidth}{!}{\textbf{Ours} (Non Blind)}}
\end{minipage}\hspace{-0.07cm}

}\vspace{-0.3cm}
\caption{Example of image denoising of a DnD image. Results of the proposed method shown when ResNet~\cite{he2016deep} used as backbone network.}
\label{Fig:DnD_Visual_Results}
\end{figure*}

The evaluation of DnD is separated in two categories: algorithms that use linear Bayer sensor readings or algorithms that use bilinearly demosaiced sRGB images as input. Thus, PSNR and SSIM for each technique are reported for both categories. The quantitative results with respect to prior work of our method and its ablations are shown in Fig.~\ref{fig:MultiNoise_and_DnD_Results}(b). For algorithms which have been evaluated with and without a variance stabilizing transformation (VST), the version which performs better is reported. The evaluation of algorithms that only operate on sRGB inputs is also reported. The proposed idea was tested for both categories. A blind and a non-blind version of our method had been tested for each category based on the info that $\bm{c}$ represents. The blind version uses no extra conditional information along with the noisy input image (empty $\bm{c}$). As described in Section~\ref{ssec:dense_our_subsp_rocgan}, in the non-blind version, $\bm{c}$ could contain information regarding the camera noise model (signal-dependent noise variance) and/or the camera id. As a backbone network for blind image denoising, two variants used: a) the standard ResNet and b) the best method in the literature named UPI~\cite{Brooks_CVPR2019}. In the case of RAW image domain, the first variant produced significantly higher PSNR (+1.01dB) and SSIM than UPI, while the second one impressively boosted the performance of UPI by 1.16dB. In the case of sRGB image domain, the first variant produced significantly higher PSNR (+1.15dB) and SSIM than UPI, while the second one impressively boosted the performance of UPI by 1.24dB. As a backbone network for non-blind image denoising, only the standard ResNet used since the best methods in the literature are not deep learning techniques. In the case of image RAW domain, our system produced significantly higher PSNR (+3.76dB) and SSIM compared to the second best method named BM3D~\cite{dabov2007image}+VST. In case of sRGB image domain, the improvement over BM3D+VST was 4.25dB. Also, runtimes (mean over 100 runs) reported in the literature are presented as well in Fig.~\ref{fig:MultiNoise_and_DnD_Results}(b). The runtime (excluding data transferring to GPU) of our blind model with standard ResNet as backbone network is 52ms while for the non-blind one is 63ms given as input 512$\times$512 images. Some qualitative results are given in Fig.~\ref{Fig:DnD_Visual_Results}.

\begin{figure*}[!t]
\setlength{\abovecaptionskip}{0.2cm}

\centering
\subfloat{
\begin{minipage}[t]{0.195\textwidth}
\centering
\includegraphics[width=0.9\textwidth]{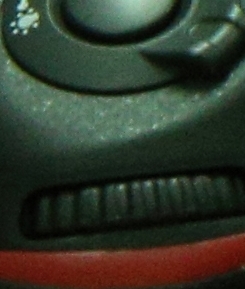}
{\footnotesize  (a) \resizebox{0.55\linewidth}{!}{Noisy image}}
\end{minipage}\hspace{-0.07cm}

\begin{minipage}[t]{0.195\textwidth}
\centering
\includegraphics[width=0.9\textwidth]{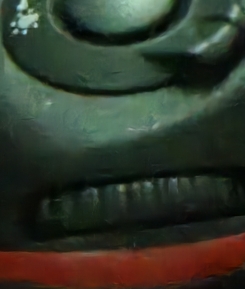}
{\footnotesize  (b) \resizebox{0.65\linewidth}{!}{RoCGAN~\cite{chrysos2019}}}
\end{minipage}\hspace{-0.07cm}

\begin{minipage}[t]{0.195\textwidth}
\centering
\includegraphics[width=0.9\textwidth]{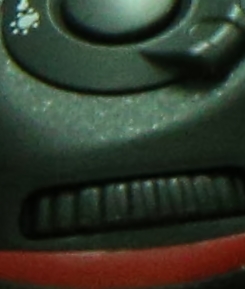}
{\footnotesize  (c) \resizebox{0.5\linewidth}{!}{BM3D~\cite{dabov2007image}}}
\end{minipage}\hspace{-0.07cm}

\begin{minipage}[t]{0.195\textwidth}
\centering
\includegraphics[width=0.9\textwidth]{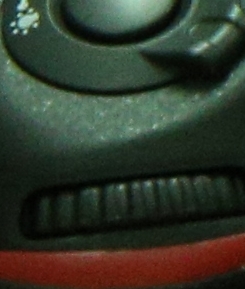}
{\footnotesize  (d) \resizebox{0.6\linewidth}{!}{DnCNN-B~\cite{zhang2017beyond}}}
\end{minipage}\hspace{-0.07cm}

\begin{minipage}[t]{0.195\textwidth}
\centering
\includegraphics[width=0.9\textwidth]{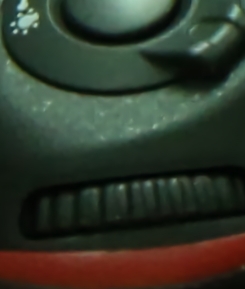}
{\footnotesize  (e) \resizebox{0.7\linewidth}{!}{CBDNet-JPEG~\cite{Guo2019Cbdnet}}}
\end{minipage}\hspace{-0.07cm}
}

\subfloat{
\begin{minipage}[t]{0.195\textwidth}
\centering
\includegraphics[width=0.9\textwidth]{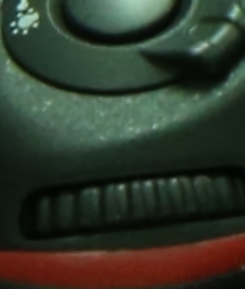}
{\footnotesize  (f) \resizebox{0.7\linewidth}{!}{RIDNet~\cite{Anwar2019RIDNET}}}
\end{minipage}\hspace{-0.07cm}

\begin{minipage}[t]{0.195\textwidth}
\centering
\includegraphics[width=0.9\textwidth]{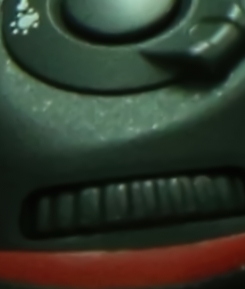}
{\footnotesize  (g) \resizebox{0.4\linewidth}{!}{NC~\cite{NC_2015}}}
\end{minipage}\hspace{-0.07cm}

\subfloat{
\begin{minipage}[t]{0.195\textwidth}
\centering
\includegraphics[width=0.9\textwidth]{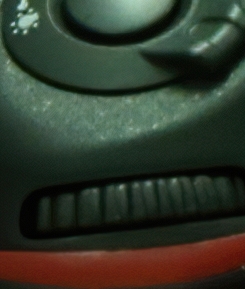}
{\footnotesize  (h) \resizebox{0.6\linewidth}{!}{\textbf{Ours} (Blind)}}    \end{minipage}\hspace{-0.07cm}
}

\begin{minipage}[t]{0.195\textwidth}
\centering
\includegraphics[width=0.9\textwidth]{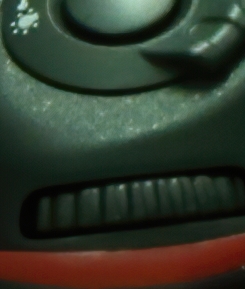}
{\footnotesize  (g) \resizebox{0.7\linewidth}{!}{\textbf{Ours} (Non Blind)}}
\end{minipage}\hspace{-0.08cm}

}\vspace{-0.3cm}
\caption{Example of image denoising of a Nam image. Results of the proposed method shown when ResNet~\cite{he2016deep} used as backbone network.}
\label{Fig:Nam_Visual_Results_2}
\end{figure*}

\textbf{Evaluation on Nam}: The Nam dataset consists of 11 static scenes captured by 3 consumer cameras. For each scene, 500 JPEG noisy temporal images were captured to compute the temporal nearly noise-free mean image and covariance matrix for each pixel. The quantitative results with respect to prior work are shown in Fig.~\ref{fig:Nam_and_SIDD_Results}(a). Both the blind and non-blind version of our method are evaluated. As a backbone network for blind image denoising, two variants used: a) the standard ResNet and b) the best method in the literature is named CBDNet~\cite{Guo2019Cbdnet}. The first variant produced significantly higher PSNR (+1.03dB) and SSIM than CBDNet, while the second one impressively boosted the performance of CBDNet by 1.18dB. CBDNet-JPEG~\cite{Guo2019Cbdnet} is a version of CBDNet which specifically deals with the JPEG compression. For fair comparison, we have retrained both variants by adopting this data augmentation technique. In that case, the first variant produced significantly higher PSNR (+0.89dB) and SSIM than CBDNet-JPEG, while the second one impressively boosted the performance of CBDNet-JPEG by 1.07dB. As a backbone network for non-blind image denoising, only the standard ResNet used since the best methods in bibliography are not deep learning techniques. Our system produced significantly higher PSNR (+0.97dB) and SSIM compared to the second best method named WNNM~\cite{Gu2014WNN}. Some qualitative results are given in Fig.~\ref{Fig:Nam_Visual_Results_2}.

\newcommand{\Nam}{
\resizebox{0.55\linewidth}{!}{$
\begin{tabular}{ l || l || l || l }
\multicolumn{1}{c||}{} & 
\multicolumn{1}{c||}{} & 
\multicolumn{1}{c||}{} & 
\multicolumn{1}{c}{} \\
\textbf{Method}       & \textbf{Type}     &  \textbf{PSNR}      & \textbf{SSIM}   \\
\hline
CDnCNN-B~\cite{zhang2017beyond} & Blind & 37.49 & 0.9272 \\
TWSC~\cite{xu2018TWSC}                           & Blind & 37.52  & 0.9292 \\
MCWNNM~\cite{Gu2014WNN}  & Blind & 37.91  & 0.9322 \\
RoCGAN~\cite{chrysos2019} & Blind & 38.52 & 0.9517 \\
Whitenner~\cite{WhiteNNer_2019} & Blind & 38.62  & 0.9527 \\
RIDNet~\cite{Anwar2019RIDNET} & Blind & 39.09  & 0.9591 \\
BM3D~\cite{dabov2007image} & Non-blind  & 39.84 & 0.9657 \\
\textbf{CBDNet}~\cite{Guo2019Cbdnet} & Blind & 40.02 & 0.9687 \\
NC~\cite{NC_2015}        & \textcolor{blue}{Blind} & \textcolor{blue}{40.41}  & \textcolor{blue}{0.9731} \\
WNNM~\cite{Gu2014WNN}  & \textcolor{purple}{Non-blind} & \textcolor{purple}{41.04} & \textcolor{purple}{0.9768} \\
\hdashline 
\textbf{Ours} (empty $\bm{c}$, \textbf{ResNet}~\cite{he2016deep}) & Blind & 41.05 (+1.03) & 0.9772 \\
\textbf{Ours} (empty $\bm{c}$, \textbf{CBDNet}) & \textcolor{blue}{Blind} & \textcolor{blue}{41.20} (\textcolor{red}{+1.18}) & \textcolor{blue}{0.9783} \\
\textbf{Ours} (\emph{Non} empty $\bm{c}$, \textbf{ResNet}) & \textcolor{purple}{Non-Blind} & \textcolor{purple}{42.01} (\textcolor{red}{+0.97}) & \textcolor{purple}{0.9830}\\
\midrule[\heavyrulewidth]
\midrule[\heavyrulewidth]
\textcolor{orange}{CBDNet-JPEG}~\cite{Guo2019Cbdnet} & \textcolor{orange}{Blind} & \textcolor{orange}{41.31} & \textcolor{orange}{0.9784} \\
\hdashline 
\textbf{Ours} (empty $\bm{c}$, \textbf{ResNet}) & Blind & 42.20 (+0.89) & 0.9855 \\
\textbf{Ours} (empty $\bm{c}$, \textcolor{orange}{\textbf{CBDNet}-JPEG}) & \textcolor{orange}{Blind} & \textcolor{orange}{42.38} (\textcolor{red}{+1.07}) & \textcolor{orange}{0.9867} \\
\toprule
\end{tabular}$}
}

\newcommand{\SIDD}{
\resizebox{0.47\linewidth}{!}{$
\begin{tabular}{ l || l || l }
\multicolumn{1}{c||}{} & 
\multicolumn{1}{c||}{} & 
\multicolumn{1}{c}{} \\
\textbf{Method}       & \textbf{Type}     &  \textbf{PSNR}   \\
\hline
DnCNN-B~\cite{zhang2017beyond} & Blind & 26.21 \\
FFDNet~\cite{2017FFDNetTA} & Blind & 29.20 \\
CBDNet-JPEG~\cite{Guo2019Cbdnet} & Blind & 30.78 \\
BM3D~\cite{dabov2007image} & \textcolor{purple}{Non-blind}  & \textcolor{purple}{30.88}\\
Whitenner~\cite{WhiteNNer_2019} & Blind & 37.57 \\
RoCGAN~\cite{chrysos2019} & Blind & 37.72 \\
\textcolor{blue}{RIDNet}~\cite{Anwar2019RIDNET} & \textcolor{blue}{Blind} & \textcolor{blue}{38.71} \\
\hdashline 
\textbf{Ours} (empty $\bm{c}$, \textbf{\textcolor{blue}{RIDNet}}) & Blind & 39.82 (+1.11) \\
\textbf{Ours} (empty $\bm{c}$, \textbf{ResNet}~\cite{he2016deep}) & \textcolor{blue}{Blind} & \textcolor{blue}{39.85 (\textcolor{red}{+1.14})} \\
\textbf{Ours} (\emph{Non} empty $\bm{c}$, \textbf{ResNet}) & \textcolor{purple}{Non-Blind} &  \textcolor{purple}{39.81} (\textcolor{red}{+8.93})\\
\toprule
\end{tabular}$}
}

\begin{figure}%
  \centering
  \subfloat[][]{\Nam}%
  \subfloat[][]{\SIDD}\vspace{-0.4cm}
  \caption{(a) The quantitative results on the Nam benchmark and (b) the quantitative results on the SIDD benchmark. Regarding our method, in parentheses we define the type of denosing plus the used backbone network.\label{fig:Nam_and_SIDD_Results}}%
\end{figure}

\begin{figure*}[!t]
\setlength{\abovecaptionskip}{0.2cm}

\centering
\subfloat{
\begin{minipage}[t]{0.170\textwidth}
\centering
\includegraphics[width=1\textwidth]{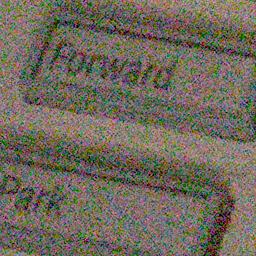}
{\footnotesize  (a) \resizebox{0.55\linewidth}{!}{Noisy image}}
\end{minipage}\hspace{-0.07cm}

\begin{minipage}[t]{0.170\textwidth}
\centering
\includegraphics[width=1\textwidth]{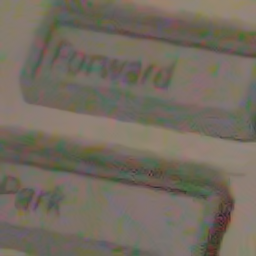}
{\footnotesize  (b) \resizebox{0.55\linewidth}{!}{RoCGAN~\cite{chrysos2019}}}
\end{minipage}\hspace{-0.07cm}

\begin{minipage}[t]{0.170\textwidth}
\centering
\includegraphics[width=1\textwidth]{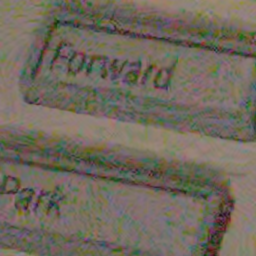}
{\footnotesize  (c) \resizebox{0.45\linewidth}{!}{BM3D~\cite{dabov2007image}}}
\end{minipage}\hspace{-0.07cm}

\begin{minipage}[t]{0.170\textwidth}
\centering
\includegraphics[width=1\textwidth]{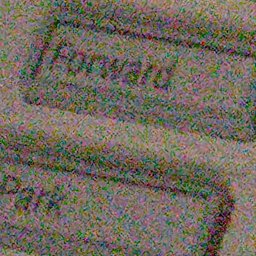}
{\footnotesize  (d) \resizebox{0.5\linewidth}{!}{FFDNet~\cite{2017FFDNetTA}}}
\end{minipage}\hspace{-0.07cm}

\begin{minipage}[t]{0.170\textwidth}
\centering
\includegraphics[width=1\textwidth]{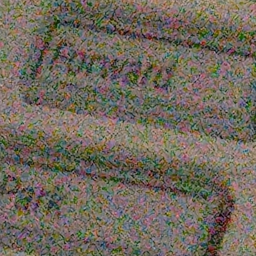}
{\footnotesize  (e) \resizebox{0.55\linewidth}{!}{DnCNN-B~\cite{zhang2017beyond}}}
\end{minipage}\hspace{-0.07cm}
}

\subfloat{
\begin{minipage}[t]{0.170\textwidth}
\centering
\includegraphics[width=1\textwidth]{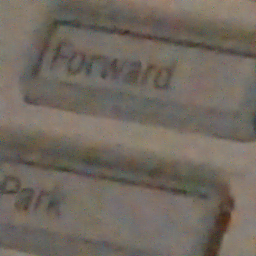}
{\footnotesize  (f) \resizebox{0.7\linewidth}{!}{CBDNet-JPEG~\cite{Guo2019Cbdnet}}}
\end{minipage}\hspace{-0.07cm}

\begin{minipage}[t]{0.170\textwidth}
\centering
\includegraphics[width=1\textwidth]{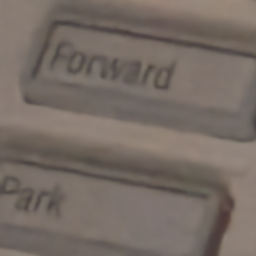}
{\footnotesize  (g) \resizebox{0.4\linewidth}{!}{RIDNet~\cite{Anwar2019RIDNET}}}
\end{minipage}\hspace{-0.07cm}

\subfloat{
\begin{minipage}[t]{0.170\textwidth}
\centering
\includegraphics[width=1\textwidth]{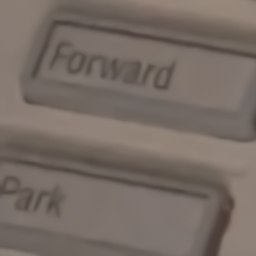}
{\footnotesize  (h) \resizebox{0.5\linewidth}{!}{\textbf{Ours} (Blind)}}    \end{minipage}\hspace{-0.07cm}
}

\begin{minipage}[t]{0.170\textwidth}
\centering
\includegraphics[width=1\textwidth]{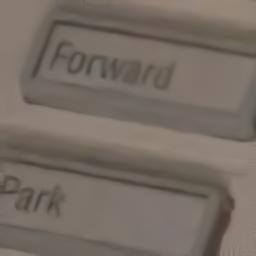}
{\footnotesize  (g) \resizebox{0.7\linewidth}{!}{\textbf{Ours} (Non Blind)}}
\end{minipage}\hspace{-0.08cm}

}\vspace{-0.3cm}
\caption{Example of image denoising of a SIDD image. Results of the proposed method shown when ResNet~\cite{he2016deep} used as backbone network.}
\label{Fig:SIDD_Visual_Results}
\end{figure*}

\textbf{Evaluation on SIDD}: SSID is real noise dataset which a large number of available test (validation) images. The quantitative results on the SIDD benchmark with respect to prior work are shown in Fig.~\ref{fig:Nam_and_SIDD_Results}(b). Both the blind and non-blind version of our method are evaluated. As a backbone network for blind image denoising, two variants used: a) the standard ResNet and b) the best method in the literature named RIDNet~\cite{Anwar2019RIDNET}. The first variant produced significantly higher PSNR (+1.11dB) than RIDNet, while the second one impressively boosted the performance of RIDNet by 1.14dB. As a backbone network for non-blind image denoising, only the standard ResNet was used since the best method in the literature, named BM3D~\cite{dabov2007image}, is not a deep learning technique. Our system produced significantly higher PSNR (+8.93dB) compared to BM3D. Some qualitative results are given in Fig.~\ref{Fig:SIDD_Visual_Results}.

Since the idea behind our method favours the multi-camera noise reduction task, there is a significant improvement in terms of performance for DnD and Nam benchmarks. Based on all our experiments, the proposed idea is general and can be appended at the end of existing image denoising methods to significantly improve their performance. In addition, the proposed idea restores better the true colors which are closer to the original pixel values than the competing methods. Also, by directly characterizing the image noise, our method avoids in great degree the image over-smoothing.

\section{Conclusions}\label{sec:Conclusions}

In this work, we show that is easier to turn noisy images into clean images only by looking at the structure of the residual image. We introduce the idea of a cGAN which explicitly leverages structure in the image noise space of the model. In that case, the generator, by adopting the residual learning, promotes the removal from the noisy image only that information which spans the manifold of the image noise. Our model significantly outperforms existing state-of-the-art architectures.

\bibliographystyle{splncs04}
\bibliography{egbib}
\end{document}